\documentclass[aps,prl,twocolumn,showpacs,superscriptaddress,nofootinbib]{revtex4}

\usepackage{amsmath}   
\usepackage{graphicx}  
\usepackage{xspace}
\usepackage{units}
\usepackage[caption=false,labelformat=simple]{subfig}
\usepackage{hyperref}

\newcommand{\minerva}{MINERvA\xspace}

\newcommand{\genie}{\textsc{genie}\xspace}
\newcommand{\neut}{\textsc{neut}\xspace}

\newcommand{\nue}{\ensuremath{\nu_{e}}\xspace}
\newcommand{\numu}{\ensuremath{\nu_{\mu}}\xspace}

\newcommand{\numubar}{\ensuremath{\bar{\nu}_{\mu}}\xspace}

\newcommand{\pip}{\ensuremath{\pi^{+}}\xspace}

\newcommand{\pim}{\ensuremath{\pi^{-}}\xspace}

\newcommand{\piz}{\ensuremath{\pi^{0}}\xspace}
\newcommand{\lam}{\ensuremath{\Lambda}\xspace}
\newcommand{\sig}{\ensuremath{\Sigma}\xspace}

\newcommand{\kp}{\ensuremath{K^{+}}\xspace}

\newcommand{\km}{\ensuremath{K^{-}}\xspace}

\newcommand{\kz}{\ensuremath{K^{0}}\xspace}

\newcommand{\kzS}{\ensuremath{K^{0}_{\mathrm{S}}}\xspace}

\newcommand{\mup}{\ensuremath{\mu^{+}}\xspace}

\newcommand{\pknu}{\ensuremath{p \rightarrow K^{+} \bar{\nu}}\xspace}

\newcommand{\sizecheck}{0} 
\newcommand{\PRLsupp}{0}   
\ifnum\PRLsupp=0

\else

\fi

\newif\ifpdf
\ifx\pdfoutput\undefined
   \pdffalse
\else
   \pdfoutput=1
   \pdftrue
\fi
\ifpdf
   \usepackage{graphicx}
   \usepackage{epstopdf}
\else
   \usepackage{graphicx}
\fi

\begin{document}

\title{Measurement of neutral-current $K^+$ production by neutrinos using MINERvA}



\newcommand{\deceased}{Deceased}


\newcommand{\Rutgers}{Rutgers, The State University of New Jersey, Piscataway, New Jersey 08854, USA}
\newcommand{\Hampton}{Hampton University, Dept. of Physics, Hampton, VA 23668, USA}
\newcommand{\Dortmund}{Institute of Physics, Dortmund University, 44221, Germany }
\newcommand{\Otterbein}{Department of Physics, Otterbein University, 1 South Grove Street, Westerville, OH, 43081 USA}
\newcommand{\JMU}{James Madison University, Harrisonburg, Virginia 22807, USA}
\newcommand{\Florida}{University of Florida, Department of Physics, Gainesville, FL 32611}
\newcommand{\UCIrvine}{Department of Physics and Astronomy, University of California, Irvine, Irvine, California 92697-4575, USA}
\newcommand{\CBPF}{Centro Brasileiro de Pesquisas F\'{i}sicas, Rua Dr. Xavier Sigaud 150, Urca, Rio de Janeiro, Rio de Janeiro, 22290-180, Brazil}
\newcommand{\PUCP}{Secci\'{o}n F\'{i}sica, Departamento de Ciencias, Pontificia Universidad Cat\'{o}lica del Per\'{u}, Apartado 1761, Lima, Per\'{u}}
\newcommand{\INRM}{Institute for Nuclear Research of the Russian Academy of Sciences, 117312 Moscow, Russia}
\newcommand{\Jlab}{Jefferson Lab, 12000 Jefferson Avenue, Newport News, VA 23606, USA}
\newcommand{\Pittsburgh}{Department of Physics and Astronomy, University of Pittsburgh, Pittsburgh, Pennsylvania 15260, USA}
\newcommand{\Guanajuato}{Campus Le\'{o}n y Campus Guanajuato, Universidad de Guanajuato, Lascurain de Retana No. 5, Colonia Centro, Guanajuato 36000, Guanajuato M\'{e}xico.}
\newcommand{\Athens}{Department of Physics, University of Athens, GR-15771 Athens, Greece}
\newcommand{\Tufts}{Physics Department, Tufts University, Medford, Massachusetts 02155, USA}
\newcommand{\WM}{Department of Physics, College of William \& Mary, Williamsburg, Virginia 23187, USA}
\newcommand{\FNAL}{Fermi National Accelerator Laboratory, Batavia, Illinois 60510, USA}
\newcommand{\Purdue}{Department of Chemistry and Physics, Purdue University Calumet, Hammond, Indiana 46323, USA}
\newcommand{\MCLA}{Massachusetts College of Liberal Arts, 375 Church Street, North Adams, MA 01247}
\newcommand{\UMD}{Department of Physics, University of Minnesota -- Duluth, Duluth, Minnesota 55812, USA}
\newcommand{\Northwestern}{Northwestern University, Evanston, Illinois 60208}
\newcommand{\UNI}{Universidad Nacional de Ingenier\'{i}a, Apartado 31139, Lima, Per\'{u}}
\newcommand{\Rochester}{University of Rochester, Rochester, New York 14627 USA}
\newcommand{\Austin}{Department of Physics, University of Texas, 1 University Station, Austin, Texas 78712, USA}
\newcommand{\USM}{Departamento de F\'{i}sica, Universidad T\'{e}cnica Federico Santa Mar\'{i}a, Avenida Espa\~{n}a 1680 Casilla 110-V, Valpara\'{i}so, Chile}
\newcommand{\Geneva}{University of Geneva, 1211 Geneva 4, Switzerland}
\newcommand{\Chicago}{Enrico Fermi Institute, University of Chicago, Chicago, IL 60637 USA}
\newcommand{\hired}{}
\newcommand{\OregonState}{Department of Physics, Oregon State University, Corvallis, Oregon 97331, USA}
\newcommand{\wroclaw}{plac Uniwersytecki 1, 50-137 Wrocław, Poland}
\newcommand{\oxford}{}
\newcommand{\umiss}{University of Mississippi, Oxford, Mississippi 38677, USA}
\newcommand{\upenn}{209 S. 33rd St. Philadelphia, PA 19104}
\newcommand{\bmeThanks}{now at SLAC National Accelerator Laboratory, Stanford, CA 94309, USA}
\newcommand{\chrisThanks}{now at Lawrence Berkeley National Laboratory, Berkeley, CA 94720, USA}
\newcommand{\higueraThanks}{now at University of Houston, Houston, TX 77204, USA}
\newcommand{\damartinezThanks}{now at Illinois Institute of Technology, Chicago, IL 60616, USA}
\newcommand{\joelmousseauThanks}{now at University of Michigan, Ann Arbor, MI 48109, USA}
\newcommand{\twaltonThanks}{now at Fermi National Accelerator Laboratory, Batavia, IL 60510, USA}
\newcommand{\jwolcottThanks}{now at Tufts University, Medford, MA 02155, USA}
\newcommand{\mcgivernThanks}{now at Fermi National Accelerator Laboratory, Batavia, IL 60510, USA}

\newcommand{\MichiganState}{Department of Physics, Michigan State University, East Lansing, MI, USA}

\author{C.M.~Marshall}\thanks{\chrisThanks}  \affiliation{\Rochester}



\author{L.~Aliaga}                        \affiliation{\WM}  \affiliation{\PUCP}
\author{O.~Altinok}                       \affiliation{\Tufts}
\author{L.~Bellantoni}                    \affiliation{\FNAL}
\author{A.~Bercellie}                     \affiliation{\Rochester}
\author{M.~Betancourt}                    \affiliation{\FNAL}
\author{A.~Bodek}                         \affiliation{\Rochester}
\author{A.~Bravar}                        \affiliation{\Geneva}
\author{T.~Cai}                           \affiliation{\Rochester}
\author{M.F.~Carneiro}                    \affiliation{\OregonState}
\author{H.~da~Motta}                      \affiliation{\CBPF}
\author{S.A.~Dytman}                      \affiliation{\Pittsburgh}
\author{G.A.~D\'{i}az~}                   \affiliation{\Rochester}  \affiliation{\PUCP}
\author{M.~Dunkman}                       \affiliation{\MichiganState}
\author{B.~Eberly}\thanks{\bmeThanks}     \affiliation{\Pittsburgh}
\author{E.~Endress}                       \affiliation{\PUCP}
\author{J.~Felix}                         \affiliation{\Guanajuato}
\author{L.~Fields}                        \affiliation{\FNAL}  \affiliation{\Northwestern}
\author{R.~Fine}                          \affiliation{\Rochester}
\author{A.M.~Gago}                        \affiliation{\PUCP}
\author{R.Galindo}                        \affiliation{\USM}
\author{H.~Gallagher}                     \affiliation{\Tufts}
\author{A.~Ghosh}                         \affiliation{\USM}  \affiliation{\CBPF}
\author{T.~Golan}                         \affiliation{\wroclaw}  \affiliation{\Rochester}
\author{R.~Gran}                          \affiliation{\UMD}
\author{D.A.~Harris}                      \affiliation{\FNAL}
\author{A.~Higuera}\thanks{\higueraThanks}  \affiliation{\Rochester}  \affiliation{\Guanajuato}
\author{K.~Hurtado}                       \affiliation{\CBPF}  \affiliation{\UNI}
\author{J.~Kleykamp}                      \affiliation{\Rochester}
\author{M.~Kordosky}                      \affiliation{\WM}
\author{T.~Le}                            \affiliation{\Tufts}  \affiliation{\Rutgers}
\author{E.~Maher}                         \affiliation{\MCLA}
\author{S.~Manly}                         \affiliation{\Rochester}
\author{W.A.~Mann}                        \affiliation{\Tufts}
\author{D.A.~Martinez~Caicedo}\thanks{\damartinezThanks}  \affiliation{\CBPF}
\author{K.S.~McFarland}                   \affiliation{\Rochester}  \affiliation{\FNAL}
\author{C.L.~McGivern}\thanks{\mcgivernThanks}  \affiliation{\Pittsburgh}
\author{A.M.~McGowan}                     \affiliation{\Rochester}
\author{B.~Messerly}                      \affiliation{\Pittsburgh}
\author{J.~Miller}                        \affiliation{\USM}
\author{A.~Mislivec}                      \affiliation{\Rochester}
\author{J.G.~Morf\'{i}n}                  \affiliation{\FNAL}
\author{J.~Mousseau}\thanks{\joelmousseauThanks}  \affiliation{\Florida}
\author{D.~Naples}                        \affiliation{\Pittsburgh}
\author{J.K.~Nelson}                      \affiliation{\WM}
\author{A.~Norrick}                       \affiliation{\WM}
\author{Nuruzzaman}                       \affiliation{\Rutgers}  \affiliation{\USM}
\author{V.~Paolone}                       \affiliation{\Pittsburgh}
\author{C.E.~Patrick}                     \affiliation{\Northwestern}
\author{G.N.~Perdue}                      \affiliation{\FNAL}  \affiliation{\Rochester}
\author{M.A.~Ram\'{i}rez}                 \affiliation{\Guanajuato}
\author{R.D.~Ransome}                     \affiliation{\Rutgers}
\author{H.~Ray}                           \affiliation{\Florida}
\author{L.~Ren}                           \affiliation{\Pittsburgh}
\author{D.~Rimal}                         \affiliation{\Florida}
\author{P.A.~Rodrigues}                   \affiliation{\umiss}  \affiliation{\Rochester}
\author{D.~Ruterbories}                   \affiliation{\Rochester}
\author{D.W.~Schmitz}                     \affiliation{\Chicago}  \affiliation{\FNAL}
\author{C.J.~Solano~Salinas}              \affiliation{\UNI}
\author{M.~Sultana}                       \affiliation{\Rochester}
\author{S.~S\'{a}nchez~Falero}            \affiliation{\PUCP}
\author{E.~Valencia}                      \affiliation{\WM}  \affiliation{\Guanajuato}
\author{T.~Walton}\thanks{\mcgivernThanks}  \affiliation{\Hampton}
\author{J.~Wolcott}\thanks{\jwolcottThanks}  \affiliation{\Rochester}
\author{M.Wospakrik}                      \affiliation{\Florida}
\author{B.~Yaeggy}                        \affiliation{\USM}
\author{D.~Zhang}                         \affiliation{\WM}

%

\collaboration{\minerva  Collaboration}\ \noaffiliation

\date{\today}

\begin{abstract}
Neutral-current production of $K^{+}$ by atmospheric neutrinos is a background in searches for the proton decay $p \rightarrow K^{+} \bar{\nu}$. Reactions such as $\nu p \rightarrow \nu K^{+} \Lambda$ are indistinguishable from proton decays when the decay products of the $\Lambda$ are below detection threshold. Events with $K^{+}$ are identified in MINERvA by reconstructing the timing signature of a $K^{+}$ decay at rest. A sample of 201 neutrino-induced neutral-current $K^{+}$ events is used to measure differential cross sections with respect to the $K^{+}$ kinetic energy, and the non-$K^{+}$ hadronic visible energy. An excess of events at low hadronic visible energy is observed relative to the prediction of the \textsc{neut}\xspace event generator. Good agreement is observed with the cross section prediction of the \textsc{genie}\xspace generator. A search for photons from $\pi^{0}$ decay, which would veto a neutral-current $K^{+}$ event in a proton decay search, is performed, and a 2$\sigma$ deficit of detached photons is observed relative to the \textsc{genie}\xspace prediction.
\end{abstract}
\ifnum\sizecheck=0  
\maketitle
\fi

Proton decay is predicted by Grand Unification Theories (GUTs)~\cite{gut1, gut2, gut3, gut4, gut5, susygut1, susygut2, susygut3, susygut4, susygut5, susygut6, susygut7, susygut8}. Models that incorporate supersymmetry predict the dominant channel to be \pknu~\cite{susygut1, susygut2, susygut3, susygut4, susygut5, susygut6, susygut7, susygut8}. Some models~\cite{susygut3, susygut4, susygut5, susygut6, susygut7, susygut8} predict proton lifetimes greater than $10^{34}$ years, consistent with the current 90\% confidence experimental bound of $5.6 \times 10^{33}$ years by Super-Kamiokande~\cite{skpknu}.

In Super-K~\cite{sknim}, the \kp from \pknu is below the Cherenkov threshold of $T_{K} = 252$~MeV, where $T_K$ is the \kp kinetic energy. Events are selected by reconstructing the \kp decay products using three techniques as described in Ref.~\cite{skpknu}. In the analysis with the lowest predicted background rate, a photon from the de-excitation of the residual nucleus is required, followed in time by a \mup from the decay $\kp \rightarrow \mup \numu$, then a ``Michel'' electron from $\mup \rightarrow e^{+} \nue \numubar$, and no other particles. Neutral-current (NC) production of \kp by atmospheric neutrinos produces the same signature when no final-state particles are above Cherenkov threshold. For example, $\nu O^{16} \rightarrow \nu K^{+} \Lambda N^{15}$ followed by $\Lambda \rightarrow p \pim$ and $\kp \rightarrow \mup \numu$ is indistinguishable from proton decay when the outgoing \kp, proton, and \pim are below threshold, and the \pim captures. 

With 40\% photo-coverage, the Super-K background prediction based on the Honda atmospheric flux~\cite{honda15} and the \neut~\cite{neut} neutrino interaction generator is 1.1 events per Megaton-year, of which NC \kp production is the largest single source at 48\% of the total background. In a 260 kiloton-year exposure, Super-K observed zero events with a predicted background of 0.38 events~\cite{skpknu}. The proposed Hyper-K~\cite{hyperk} will have an exposure of several Megaton-years, in which several background events from NC \kp production by atmospheric neutrinos would be expected. Constraining this background with a measurement of the cross section for such a process is an important input to future proton decay searches.

DUNE also plans to search for \pknu~\cite{dune}. With very low thresholds for charged hadrons, DUNE can potentially reach lower background rates by detecting the associated antikaon or hyperon present in all NC \kp reactions. However, low-energy hadrons could be present in proton decay signal events due to interactions between the \kp and the residual nucleus. The event selection must tolerate some nuclear activity to achieve high efficiency, and NC \kp production with very low visible energy can mimic the signal process. The predicted background rate comes from a single event in a 1 Megaton-year simulation with a \kp, no other tracks or $\pi^{0}$s and less than 800 MeV of total visible energy~\cite{bueno}.


Strange particle production by neutrinos has previously been observed in bubble chambers~\cite{bebc1, bebc2, bebc3, bebc4, fnal15a, fnal15b, anl2, anl12ft, bnl7ft, fnal15c, gargamelle} and in NOMAD~\cite{nomad}. Gargamelle observed three NC \kp production events~\cite{gargamelle}, and reported a neutral- to charged-current ratio for strangeness production.

In this Letter, we report measurements of NC \kp production by neutrinos in \minerva with a sample of 201 events after background subtraction. We report differential cross sections with respect to the \kp kinetic energy, and with respect to a measure of the non-\kp visible energy. To address potential proton decay backgrounds for water Cherenkov detectors, we also quantify the rate of Michel electrons and detached photons in our data relative to the prediction of \genie, the event generator used in \minerva, DUNE, and many other experiments. These measurements serve as benchmarks to cross-section predictions used in background simulations for proton decay experiments.

\minerva is a dedicated neutrino-nucleus cross section experiment in the NuMI beamline~\cite{numi} at Fermilab. The data used in this Letter were taken in a $\nu_{\mu}$-enriched beam with a peak neutrino energy of 3.5~GeV between March 2010 and April 2012, corresponding to $3.51 \times 10^{20}$ protons on target. A Geant4-based model is used to simulate the neutrino beam, and is described in Refs.~\cite{fluxpaper, leothesis}. This model is constrained to reproduce thin-target hadron production measurements on carbon~\cite{na49, na49b, mipp, kpiratio, barton}. 

The \minerva detector consists of a core of scintillator strips surrounded by electromagnetic and hadronic calorimeters. For this result, the interaction vertex is constrained to be within the central 108 planes of the scintillator tracking region and no closer than 22 cm to any edge of the planes. The fiducial volume is 5.57 metric tons, consisting of 95\% CH and 5\% other materials by mass. The design, calibration, and performance of the \minerva detector are described in Ref.~\cite{minerva_nim}.

Neutrino interactions are simulated using the \genie 2.8.4 neutrino event generator~\cite{genie}. 
Strange particles are produced primarily from hadronization of deep inelastic scattering events, and also in the decays of $\Delta$ and $N$ baryon resonances, for example $\Delta$(1750)$\rightarrow K \Sigma$. Inelastic reactions for final-state hadronic invariant mass $W < 1.7$~GeV are simulated with a tuned model of discrete baryon resonance production~\cite{reinsehgal}. Strange baryon resonances are not simulated. The transition to deep
inelastic scattering is simulated using the Bodek-Yang model~\cite{bodekyang}. Hadronization at higher energies is simulated with the AGKY model~\cite{agky} based on the gradual transition from KNO scaling~\cite{kno} to the Lund string model of PYTHIA (version 6)~\cite{pythia} with increasing $W$. Parameters that control the rate of strange particle production in hadronization are tuned such that rates of \lam and \kzS production on deuterium and neon agree with BEBC~\cite{bebc1, bebc2, bebc3, bebc4} and Fermilab 15'~\cite{fnal15a, fnal15b} bubble chamber measurements. FSI are modeled using the
INTRANUKE package~\cite{genie, dytmanFSI}. FSI for \kp in \genie 2.8.4 include elastic and inelastic scattering but not charge exchange. Pion-induced processes that produce kaons, for example $\pip n \rightarrow \kp \lam$, are not included, nor is any FSI for \lam and \sig hyperons.

In the \neut generator used by Super-K, hadronization is simulated using PYTHIA for $W > 2$~GeV, and an exclusive $\kp \lam$ process is simulated for $W < 2$~GeV~\cite{hayato}.

We define the signal process as a neutral-current reaction with at least one \kp exiting the nucleus in which the neutrino interaction occurred, with kinetic energy less than 600 MeV. There is no requirement on the neutrino flavor or helicity.

\begin{figure}
\centering
\includegraphics[width=\columnwidth]{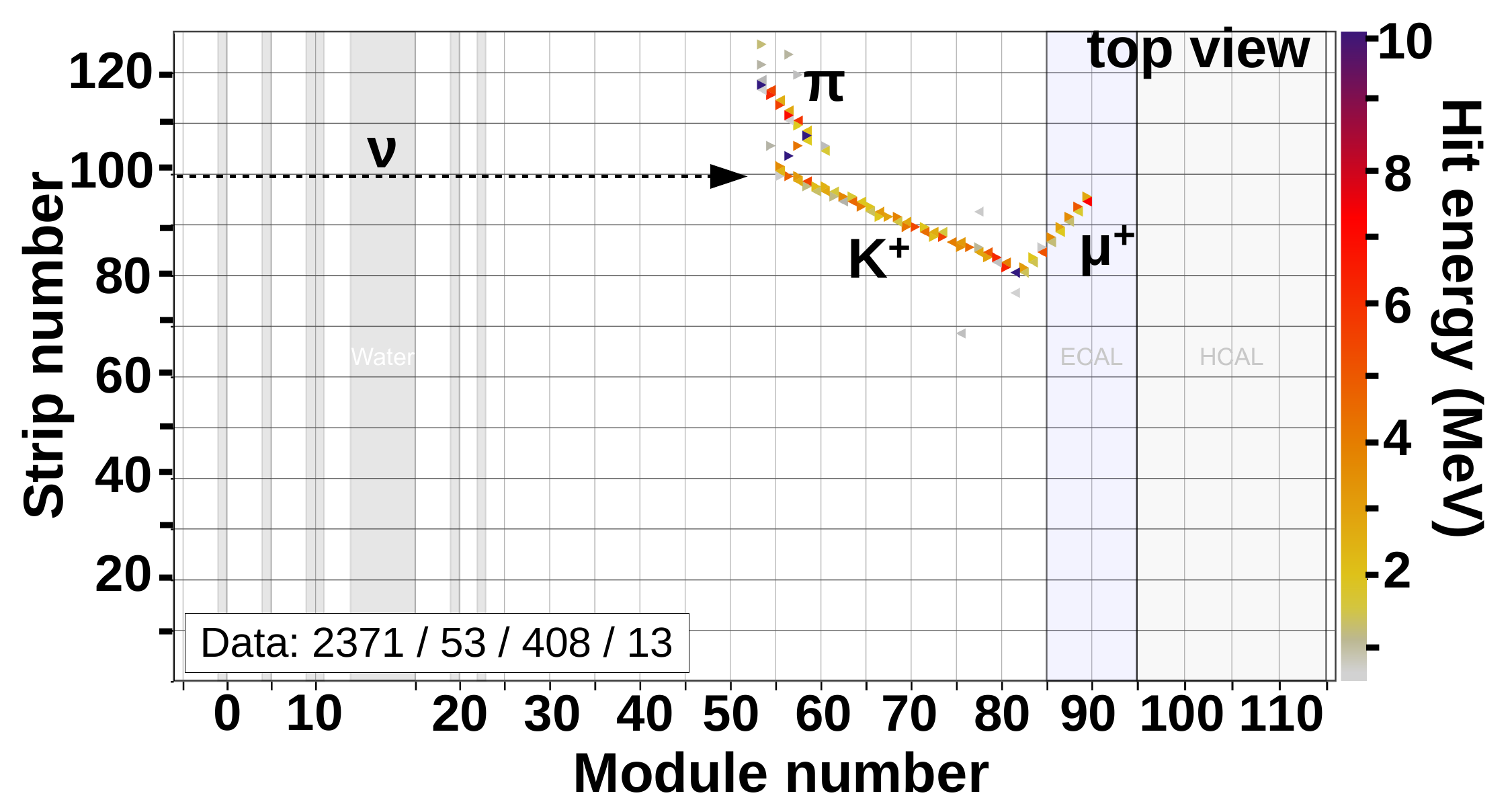}
    \vspace{-7pt}
\caption{A neutral-current \kp data candidate event in \minerva shown in the X-Z view. This event is the 13th neutrino interaction in beam spill 408, of run 2371, subrun 53. The small triangles represent time-stamped energy deposits in individual scintillator strips. A hypothesis for each particle track is given. The \mup is delayed in time relative to the \kp. The track labeled as a pion is probably from $\Sigma^- \rightarrow n \pim$ with an unobserved neutron, however it could also be a \km track. \label{fig:evtdisp}}
    \vspace{-10pt}
\end{figure}

Stopping kaons are selected by reconstructing the timing signature of a \kp decay-at-rest. The general method used to identify \kp mesons in \minerva is described in Ref.~\cite{cc_paper}. For NC events, however, a muon tag for primary vertices is not available. Consequently, the kinetic energy of candidate \kp in this analysis is required to exceed the 100 MeV tracking threshold. Kaons with kinetic energies greater than 600 MeV are not contained unless they undergo an interaction in the detector. For interacting kaons, the range-based \kp energy reconstruction underestimates the true kinetic energy because some of the \kp kinetic energy is transferred to other particles, typically nucleons. An example NC \kp candidate from data is shown in Fig.~\ref{fig:evtdisp}.

NC events with no muon at the neutrino interaction point are selected by requiring that no track other than the \kp candidate traverse more than 250 g/cm$^{2}$ of material in \minerva, where the side and downstream calorimeters are included. The efficiency of this selection for true NC events is 83\%, where the inefficiency is due to energetic non-interacting pions and protons. In simulation, 11\% of true charged-current (CC) events have muons with range less than 250 g/cm$^{2}$, which corresponds to 500 MeV of kinetic energy.

\begin{figure}
\centering
\includegraphics[width=\columnwidth]{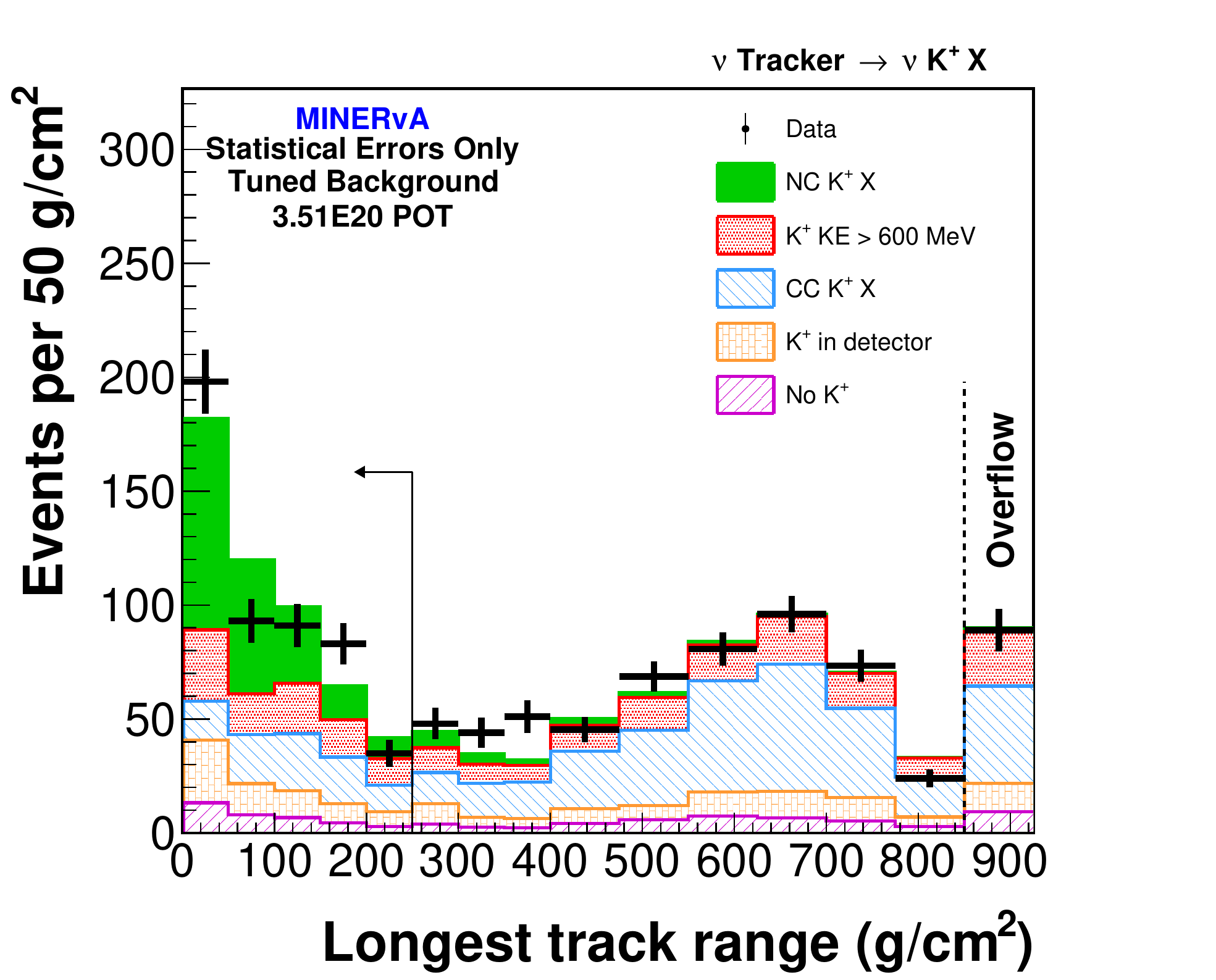}
    \vspace{-7pt}
\caption{The range of the longest reconstructed track in the event excluding the \kp, in g/cm$^{2}$. The arrow shows selected events, with events to the right forming the CC-like sideband. The highest bin includes all events above $850 g/cm^{2}$ and is not normalized by the bin width, and the lowest bin includes events in which the \kp is the only reconstructed track. \label{fig:muonrange}}
    \vspace{-10pt}
\end{figure}

The distribution of the longest track range excluding the \kp candidate is shown in Fig.~\ref{fig:muonrange}. The arrow shows the selection at 250 g/cm$^{2}$. Backgrounds are divided into four categories described in detail below: CC or NC \kp events with $T_{K} > 600$~MeV, CC \kp events with $T_{K} < 600$~MeV, CC or NC events where a \kp does not exit the struck nucleus but is produced inside the detector, and events where some other particle is misidentified as a \kp. Backgrounds are constrained using a sideband formed from events with tracks traversing more than 250 g/cm$^{2}$ of material. Events with rear-exiting tracks are excluded from the sideband in order to make the composition of the backgrounds in the signal and sideband regions more similar, as well as to exclude events at low inelasticity $y = (E_{\nu}-E_{\mu})/E_{\nu}$, since CC backgrounds in the signal region are at high $y$. Backgrounds due to beam pile-up are constrained by the method described in Ref.~\cite{cc_paper}. All other backgrounds are constrained together using the CC-rich sideband, and a scale factor of $0.96 \pm 0.23$ is applied based on the fit. The scale factor uncertainty includes effects that are highly correlated between the signal and sideband region. The resulting uncertainty on the background-subtracted data is 10\%.

A \kp with kinetic energy greater than 600 MeV will exit the inner detector unless it interacts hadronically. Because the kaon energy reconstruction is range-based, all accepted events in this kinematic regime have poorly reconstructed energies. In both CC and NC scattering, high-energy kaons originate from the same hadronization models in \genie. We apply an uncertainty to these events equal to the difference between the spectra produced by the PYTHIA and KNO hadronization models. This results in an uncertainty of $^{+46}_{-11}$\% relative to the central value, which is a gradual transition from KNO to PYTHIA with increasing $W$ between $2.3$ and $3.0$ GeV. We treat the high-energy shape of the \kp spectrum as correlated between CC and NC events, and use the CC-rich sideband to constrain NC \kp production with \kp kinetic energy greater than 600 MeV.

The dominant background is due to true CC \kp events with muon kinetic energy less than 500 MeV. The constraint from the CC-like sideband involves extrapolating in neutrino energy, $E_{\nu}$, and $y$. Sideband events with higher muon energy preferentially come from higher $E_{\nu}$ and lower $y$ than CC events in the signal region. Uncertainties due to the modeling of the $y$ distribution in \genie, as well as uncertainties on the flux shape, enter the analysis in the background subtraction~\cite{thesis}.

Another important background is due to \kp produced by hadronic interactions outside the struck nucleus, labeled ``\kp in detector'' in Fig.~\ref{fig:muonrange}. These interactions include pion reactions like $\pi^{+} n \rightarrow K^{+} \Lambda$, and \kz charge exchange $\kz p \rightarrow \kp n$. This background is present in both CC- and NC-rich samples. The largest uncertainty on the rate of these backgrounds is due to the cross section for the hadronic processes in the detector, which is correlated between the CC- and NC-like regions. We place an \textit{a priori} uncertainty of 100\% on both classes of events, which are constrained in the background fit.

We report the differential cross section with respect to the \kp kinetic energy, $T_{K}$, as well as with respect to a measure of non-\kp energy, $E_{vis}$, defined as the sum of the kinetic energy of all $\pi^{\pm}$, \km, and protons, and the total energy of all photons, \piz, and \kz. The energy sum $E_{vis}$ includes the prompt decay products of \lam and \sig hyperons. The relationship between the observed energy in \minerva and the total hadronic energy, $\nu$, depends on the relative fraction of $\nu$ carried by different particle species. The \minerva detector responds differently to hadronic showers induced by $p/\pi$ and electromagnetic showers induced by $\piz \rightarrow \gamma \gamma$. Neutrons are detected only when they scatter inside the detector and produce charged particles.

The flux-integrated differential cross section per nucleon in bin $i$ is

\begin{equation} \label{eq:diffxs}
\left(\frac{d\sigma}{dX}\right)_{i} = \frac{\sum_{j}U_{ij}\left(N_{j} - N_{j}^{bg}\right)}{\epsilon_{i}N_{nuc}\Phi\Delta_{i}},
\end{equation}

\noindent
where $j$ is the index of a reconstructed bin of variable $X = T_{K}, E_{vis}$, $U_{ij}$ is the unsmearing matrix, $N_{j}$ is the number of selected events, $N_{j}^{bg}$ is the predicted number of background events, $\epsilon_{i}$ is the selection efficiency for signal events, $N_{nuc}$ is the number of nucleons in the fiducial volume, $\Phi$ is the integrated \numu flux prediction, and $\Delta_{i}$ is the width of bin $i$.

After background subtraction, including subtracting the estimate for events with \kp kinetic energy greater than 600 MeV, there are 201 signal events in data. The overall sample purity is 41.2\% when events with $T_{K} > 600$~MeV are considered a background. The data are unfolded using a Bayesian procedure with three iterations \cite{dagostini} described in Refs.~\cite{cc_paper,thesis}. The overall selection efficiency is 4.1\%. The largest contribution to the inefficiency is the requirement that the \kp be at rest for of order 10 ns, which removes 50\% of the signal events but is necessary to obtain reasonable purity.  



\begin{figure}
\centering
\includegraphics[width=0.9\columnwidth,bb=20 20 575 520]{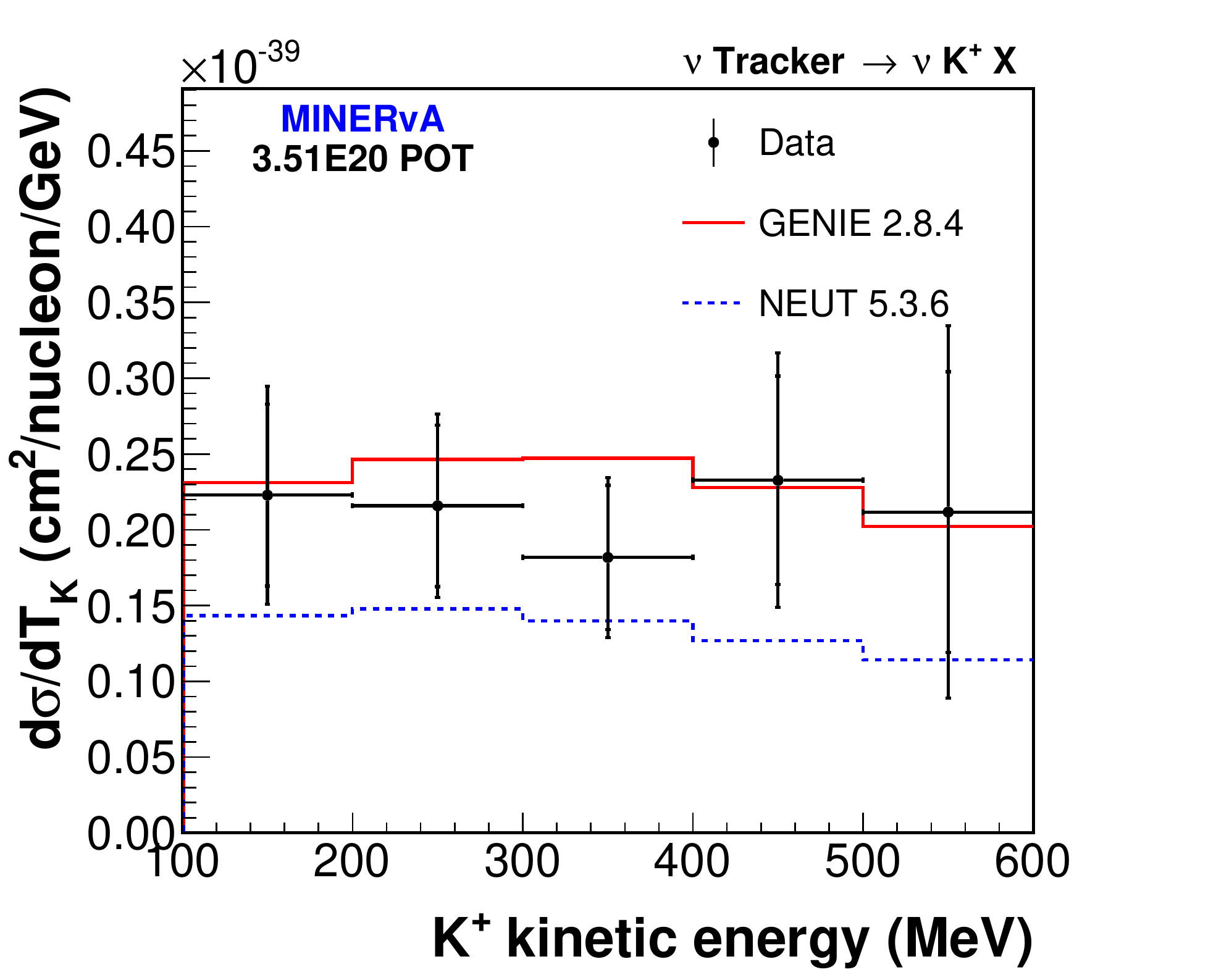}
\caption{The differential cross section as a function of \kp kinetic energy $T_K$, compared to \genie and \neut. The inner (outer) error bars represent the statistical (total) uncertainty. \label{fig:tkxs}}
\end{figure}

The differential cross section with respect to the kaon kinetic energy, $T_{K}$, is shown in Fig.~\ref{fig:tkxs}, along with predictions from \genie 2.8.4 and \neut 5.3.6. We observe a flat shape in $d\sigma/dT_{K}$ between 100 and 600 MeV, consistent with the \genie and \neut predictions. The total (shape-only) $\chi^2$ with 5 (4) degrees of freedom is 1.80 (1.05) for \genie and 6.26 (0.93) for \neut. This excess with respect to \neut is not statistically significant. Absolute and fractional uncertainties, as well as covariance matrices, are given in the supplementary material. 

\begin{figure}
\centering
\includegraphics[width=0.9\columnwidth,bb=20 20 575 554]{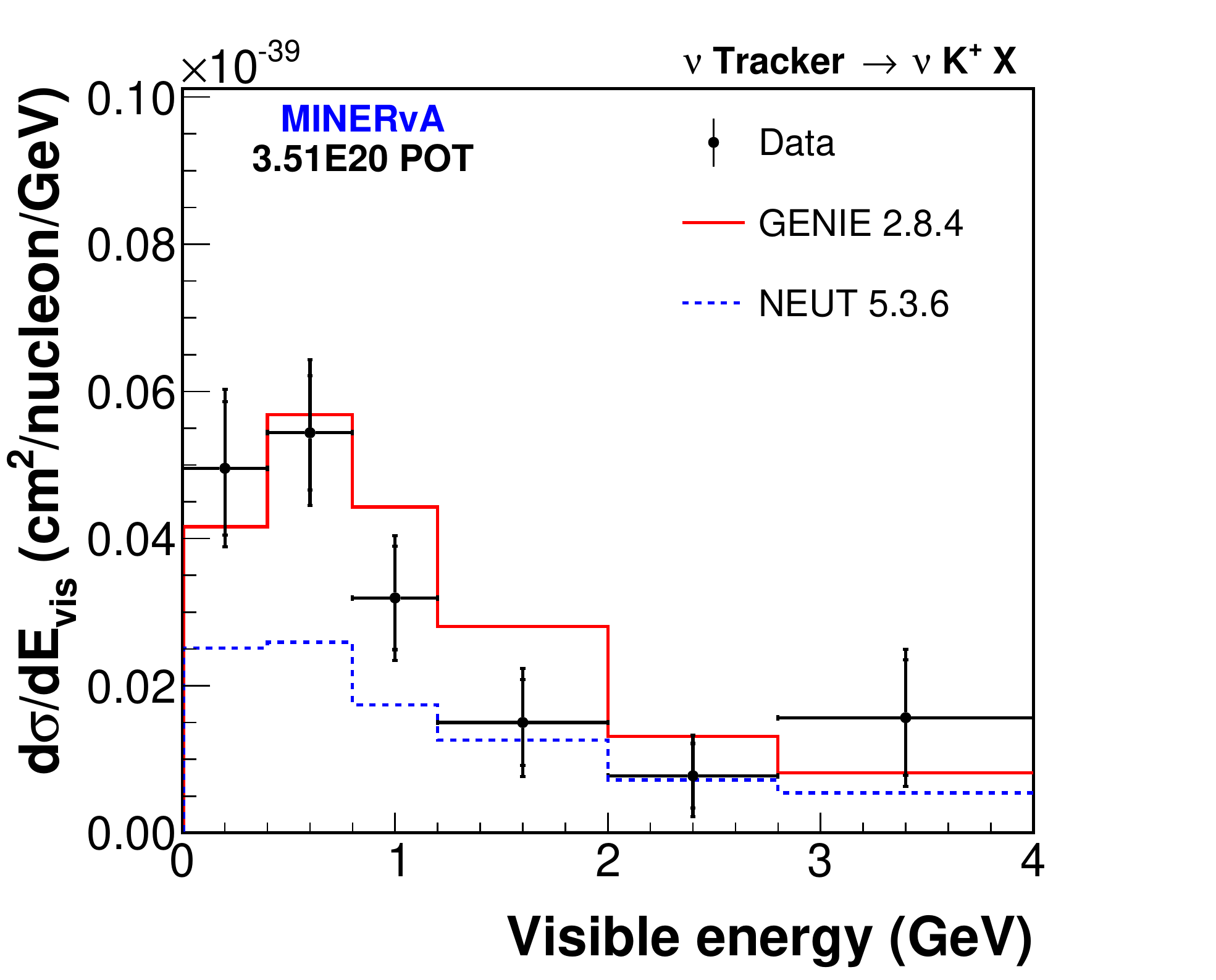}
\caption{The differential cross section as a function of non-\kp visible energy $E_{vis}$, as defined in the text, compared to \genie and \neut. The inner (outer) error bars represent the statistical (total) uncertainty. \label{fig:nuxs}}
\end{figure}

Events with a single \kp track and very little other energy are potential backgrounds in \pknu searches. The differential cross section with respect to non-kaon visible energy, $E_{vis}$, is shown in Fig.~\ref{fig:nuxs}. The total (shape-only) $\chi^2$ with 6 (5) degrees of freedom is 7.31 (6.10) for \genie and 12.13 (3.42) for \neut. The absolute $\chi^2$ for \neut in the lowest two bins is 10.48 with 2 degrees of freedom. Strangeness production in \neut is insufficient at low $W$, which is related to the final state hadronic energy, $\nu$, by $W^{2} = M^{2} + 2 M \nu - Q^{2}$, where $M$ is the nucleon mass and $Q^{2}$ is the squared four-momentum transfer to the nuclear system. $E_{vis}$ is approximately equal to $\nu - E_{K}$, where $E_{K}$ is the \kp total energy, and the difference is due to the exclusion of neutron kinetic energy and charged pion rest masses from $E_{vis}$. Absolute and fractional uncertainties, as well as covariance matrices, are given in the supplementary material. 

\neut is used by Super-K to predict the atmospheric neutrino background in the search for \pknu~\cite{skpknu}. Neutrino events satisfying the \pknu selection are expected to be at $E_{vis} < 0.8$~GeV, where the excess in our data compared to \neut is greatest. A $p \pim$ from \lam decay can have up to 630 MeV of visible energy with both particles below Cherenkov threshold. However, Super-K is a zero-threshold detector of \pip and \piz because of the observable decay products, $\pip \rightarrow \mup \rightarrow e^{+}$ and $\piz \rightarrow \gamma \gamma$. 

In NC \kp events with $E_{vis} < 0.8$~GeV, a final-state \pip is predicted in 25\% of events in \genie compared to 4\% in \neut, while a final-state \piz is predicted in 35\% of events in \genie and 39\% in \neut. These percentages include pions from prompt hyperon decays. In \neut, NC \kp production for $W < 2$~GeV is simulated as a $\kp \lam$ exclusive process without additional mesons~\cite{hayato}. This process produces $\pi^{0}$s from $\lam \rightarrow n \piz$, but never produces \pip. In \neut, the threshold for $\Sigma$ baryon associated production, or for \kp associated production with additional pions, is $W = 2$~GeV. In \genie, reactions like $\nu p \rightarrow \nu \kp \Sigma^{+} \pi^{-}$ or $\nu p \rightarrow \nu \kp \Sigma^{-} \pip$ turn on at threshold around $W = 1.8$~GeV. At low $W$, \neut is also missing reactions like $\nu n \rightarrow \nu \kp \Sigma^{-}$ as well as kaon-antikaon associated production, which could be important sources of background to \pknu due to the lack of \piz and \pip.

Estimates of \pip and \piz in $E_{vis} < 0.8$~GeV events in \minerva data are obtained from searches for Michel electrons, and for detached photons. Michel electrons are identified in \minerva by searching for activity near the endpoint of a charged particle track that is delayed in time and consistent in energy with an electron from $\mu \rightarrow e \nu \nu$. Two selection criteria are used. The first is less stringent, with a 28\% efficiency to find a Michel in an event with a \pip, and a 5.5\% fake rate. A second, more stringent, selection requires delayed activity in all three \minerva plane orientations. This reduces the rate of false positives to 1.4\%, and reduces the efficiency to 12\%.

Good agreement with \genie in the Michel electron rate is observed. With the aggressive selection, a Michel is identified in 12.7\% of simulated selected events with $E_{vis} < 0.8$~GeV, compared to ($11.8 \pm 2.6$)\% in data. With the more stringent selection, the rate is 3.8\% in simulation and ($4.0 \pm 1.5$)\% in data. Our data are not directly comparable to \neut because this comparison uses selected events and reconstructed quantities.

To search for detached photons, a visual scan was performed by three scanners on events with reconstructed $E_{vis} < 0.8$~GeV in data and simulation. Photons are expected to arise primarily from $\piz \rightarrow \gamma \gamma$, with \piz from $\lam \rightarrow n \piz$, but could also be due to $\Sigma^{0} \rightarrow \lam \gamma$. A photon was identified in 23\% of data events compared to 33\% of simulated events, giving a ratio of data to simulation of $0.70 ^{+0.15}_{-0.13}$ (stat.) $\pm 0.01$ (syst.), where the statistical uncertainty is due to the finite scanning samples of 177 events each in data and simulation, and the systematic uncertainty is the disagreement between scanners. We observe an indication of an overprediction of the photon rate in NC \kp events that is significant at $2\sigma$.

In neutral-current scattering, a \kp is always produced in association with a hyperon or antikaon. While \minerva does not reconstruct individual hyperons, the pion content of the final state gives some sensitivity to different final states. For example, a \piz was identified in 40\% of $\kp \lam$ simulated events, compared to 18\% of $\kp \km$ events. Our data prefer a reduction in the $\kp \lam$ rate and a corresponding enhancement of $\kp \km$, which would reduce the \piz rate while preserving the total \kp cross section.

The statistics obtained by \minerva in this channel are far larger than what could be achieved with a sensitivte detector exposed to atmospheric neutrinos. The expected number of NC \kp interactions with $E_{vis} < 0.8$~GeV on a $CH$ target is computed using \genie~\cite{genie} for the \minerva flux~\cite{fluxpaper} and the atmospheric neutrino flux at Kamioka~\cite{honda15}. For this signal, the \minerva low-energy run of $3.51 \times 10^{20}$ protons on target and a 5.57-ton fiducial volume is equivalent to a 58.4 Megaton-year exposure to atmospheric neutrinos, and a 9.6 Megaton-year exposure to atmospheric antineutrinos.

In conclusion, we report the first high-statistics measurement of NC \kp production by neutrinos. The measured cross section is in disagreement with the \neut prediction at low $E_{vis}$, and in good agreement with the \genie prediction. Searches were performed for Michel electrons and for photon conversions, which would cause an event to be rejected in a proton decay search. We observe good agreement with \genie in the rate of Michel electrons in low-$E_{vis}$ events, and a deficit of 30\% in the number of photons compared to \genie. Our data indicate that the \neut cross section for NC \kp production is too small by a factor of two in the kinematic region of most importance to searches for \pknu.

\begin{acknowledgments}

We thank Yoshinari Hayato for his input regarding the \neut generator. We thank Callum Wilkinson and Patrick Stowell for preparing the \neut generator predictions. This work was supported by the Fermi National Accelerator Laboratory under US Department of Energy contract No. DE-AC02-07CH11359 which included the MINERvA construction project. Construction support was also granted by the United States National Science Foundation under Award PHY-0619727 and by the University of Rochester. Support for participating scientists was provided by NSF and DOE (USA), by CAPES and CNPq (Brazil), by CoNaCyT (Mexico), by CONICYT programs including FONDECYT (Chile), by CONCYTEC, DGI-PUCP and IDI/IGI-UNI (Peru), and by Latin American Center for Physics (CLAF). We thank the MINOS Collaboration for use of its near detector data. We acknowledge the dedicated work of the Fermilab staff responsible for the operation and maintenance of the beamline and detector, and the Fermilab Computing Division for support of data processing.

\end{acknowledgments}

\end{document}


\thispagestyle{empty}

\newcommand{\qsq}{\ensuremath{Q^2_{QE}}\xspace}
\renewcommand{\textfraction}{0.05}
\renewcommand{\topfraction}{0.95}
\renewcommand{\bottomfraction}{0.95}
\renewcommand{\floatpagefraction}{0.95}
\renewcommand{\dblfloatpagefraction}{0.95}
\renewcommand{\dbltopfraction}{0.95}
\setcounter{totalnumber}{5}
\setcounter{bottomnumber}{3}
\setcounter{topnumber}{3}
\setcounter{dbltopnumber}{3}

\appendix{Supplementary Material}\hfill\vspace*{4ex}\

The supplementary material contains cross section tables and covariance matrices. Table~\ref{tab:tkxs} is the differential cross section $d\sigma/dT_K$ with absolute statistical and systematic uncertainties. Fractional uncertainties broken down by category are given in Table~\ref{tab:tksyst}. Tables~\ref{tab:tkstatcov},~\ref{tab:tkfluxcov}, and~\ref{tab:tksyscov} are covariance matrices for the statistical, flux, and other systematics, respectively.

Table~\ref{tab:nuxs} is the differential cross section $d\sigma/dE_{vis}$ with absolute statistical and systematic uncertainties. Fractional uncertainties broken down by category are given in Table~\ref{tab:nusyst}. Tables~\ref{tab:nustatcov},~\ref{tab:nufluxcov}, and~\ref{tab:nusyscov} are covariance matrices for the statistical, flux, and other systematics, respectively.

In Tables~\ref{tab:tksyst} and~\ref{tab:nusyst}, systematic uncertainties are broken into catories. The ``Bkg. model'' category includes uncertainties on charged-current \kp production, \kp produced by strong reactions inside the detector, and charge exchange, which are constrained by the sideband fit. The energy scale uncertainty is the effect of events moving between bins due to the uncertain reconstruction of the kaon energy. The ``Kaon int.'' category includes uncertainties on \kp and \kz charge exchange in the detector, as well as the uncertainty on \kp rescattering on carbon, which leads to misreconstruction of the \kp energy. ``Sideband tune'' is the uncertainty introduced by the tuning procedure, including the statistical uncertainty on the data in the sideband. The ``Sig. model'' uncertainty is the small effect introduced by mismodeling the cross section with respect to other variables, for example \kp angle.

\begingroup
\squeezetable
\begin{table}[h]
\begin{tabular}{l|cccc}
\hline
$T_{K}$ (MeV) & $d\sigma/dT_{K}$ & Total & Statistical & Systematic \\
\hline
100 - 200 & 0.22 & 0.07 & 0.06 & 0.04 \\ 
200 - 300 & 0.21 & 0.06 & 0.05 & 0.03 \\ 
300 - 400 & 0.18 & 0.05 & 0.05 & 0.03 \\ 
400 - 500 & 0.23 & 0.08 & 0.07 & 0.05 \\ 
500 - 600 & 0.21 & 0.12 & 0.09 & 0.08 \\ 
\hline
\end{tabular}
\caption{The differential cross section as a function of \kp kinetic energy, $T_K$, for NC \kp production along with absolute statistical and systematic uncertainties, all in units of $10^{-39}$~cm$^{2}$/nucleon/GeV.}
\label{tab:tkxs}
\end{table}
\endgroup

\begingroup
\squeezetable
\begin{table}
\begin{tabular}{l|ccccc}
\hline
Source            & 100 - 200 & 200 - 300 & 300 - 400 & 400 - 500 & 500 - 600 \\ 
\hline
Statistics        & 0.27      & 0.25      & 0.26      & 0.30      & 0.44 \\ 
Bkg. model        & 0.08      & 0.06      & 0.09      & 0.14      & 0.29 \\ 
Flux              & 0.09      & 0.09      & 0.07      & 0.06      & 0.08 \\ 
Energy scale      & 0.03      & 0.01      & 0.02      & 0.02      & 0.07 \\ 
Kaon int.         & 0.13      & 0.07      & 0.02      & 0.11      & 0.10 \\ 
Sideband tune     & 0.03      & 0.03      & 0.05      & 0.05      & 0.09 \\ 
Sig. model        & 0.00      & 0.00      & 0.01      & 0.07      & 0.19 \\ 
\hline
Total             & 0.32      & 0.28      & 0.29      & 0.36      & 0.58 \\ 
\hline
\end{tabular}
\caption{Fractional statistical and systematic uncertainties reported in bins of \kp kinetic energy.}
\label{tab:tksyst}
\end{table}
\endgroup

\begingroup
\squeezetable
\begin{table}[h]
\begin{tabular}{l|ccccc}
\hline
$T_{K}$ (MeV) & 100 - 200 & 200 - 300 & 300 - 400 & 400 - 500 & 500 - 600 \\
\hline
100 - 200 & 0.358 & 0.008 & -0.023 & -0.052 & -0.030 \\
200 - 300 & 0.008 & 0.285 & -0.005 & -0.058 & -0.050 \\
300 - 400 & -0.023 & -0.005 & 0.225 & 0.011 & -0.035 \\
400 - 500 & -0.052 & -0.058 & 0.011 & 0.471 & 0.044 \\
500 - 600 & -0.030 & -0.050 & -0.035 & 0.044 & 0.855 \\
\hline
\end{tabular}
\caption{The statistical covariance introduced by the unfolding procedure on the measurement of $d\sigma/dT_{K}$, given in units of $10^{-80}$~(cm$^{2}$/nucleon/GeV)$^{2}$.}
\label{tab:tkstatcov}
\end{table}
\endgroup

\begingroup
\squeezetable
\begin{table}[h]
\begin{tabular}{l|ccccc}
\hline
$T_{K}$ (MeV) & 100 - 200 & 200 - 300 & 300 - 400 & 400 - 500 & 500 - 600 \\
\hline
100 - 200 & 0.036 & 0.033 & 0.020 & 0.024 & 0.027 \\
200 - 300 & 0.033 & 0.036 & 0.020 & 0.024 & 0.029 \\
300 - 400 & 0.020 & 0.020 & 0.014 & 0.015 & 0.017 \\
400 - 500 & 0.024 & 0.024 & 0.015 & 0.022 & 0.021 \\
500 - 600 & 0.027 & 0.029 & 0.017 & 0.021 & 0.030 \\
\hline
\end{tabular}
\caption{The covariance due to the flux uncertainty on the measurement of $d\sigma/dT_{K}$, given in units of $10^{-80}$~(cm$^{2}$/nucleon/GeV)$^{2}$.}
\label{tab:tkfluxcov}
\end{table}
\endgroup

\begingroup
\squeezetable
\begin{table}[h]
\begin{tabular}{l|ccccc}
\hline
$T_{K}$ (MeV) & 100 - 200 & 200 - 300 & 300 - 400 & 400 - 500 & 500 - 600 \\
\hline
100 - 200 & 0.133 & 0.065 & 0.039 & -0.040 & 0.029 \\
200 - 300 & 0.065 & 0.048 & 0.035 & 0.012 & 0.054 \\
300 - 400 & 0.039 & 0.035 & 0.055 & 0.070 & 0.114 \\
400 - 500 & -0.040 & 0.012 & 0.070 & 0.226 & 0.311 \\
500 - 600 & 0.029 & 0.054 & 0.114 & 0.311 & 0.615 \\
\hline
\end{tabular}
\caption{The covariance due to systematic uncertainties other than the flux, on the measurement of $d\sigma/dT_{K}$, given in units of $10^{-80}$~(cm$^{2}$/nucleon/GeV)$^{2}$.}
\label{tab:tksyscov}
\end{table}
\endgroup

\begingroup
\squeezetable
\begin{table}[h]
\begin{tabular}{l|cccc}
\hline
$E_{vis}$ (GeV) & $d\sigma/dE_{vis}$ & Total & Statistical & Systematic \\
\hline
0.0 - 0.4 & 49.5 & 10.7 & 9.1 & 5.7 \\ 
0.4 - 0.8 & 54.4 & 9.9 & 7.8 & 6.2 \\ 
0.8 - 1.2 & 31.9 & 8.4 & 7.0 & 4.8 \\ 
1.2 - 2.0 & 15.0 & 7.4 & 5.9 & 4.5 \\ 
2.0 - 2.8 & 7.8 & 5.6 & 4.4 & 3.4 \\ 
2.8 - 4.0 & 15.7 & 9.3 & 7.8 & 5.0 \\ 
\hline
\end{tabular}
\caption{The differential cross section as a function of non-\kp visible energy, $E_{vis}$, for NC \kp production with \kp kinetic energy less than 600 MeV, along with absolute statistical and systematic uncertainties, all in units of $10^{-42}$~cm$^{2}$/nucleon/GeV.}
\label{tab:nuxs}
\end{table}
\endgroup

\begingroup
\squeezetable
\begin{table}
\begin{tabular}{l|cccccc}
\hline
Source             & 0.0 - 0.4 & 0.4 - 0.8 & 0.8 - 1.2 & 1.2 - 2.0 & 2.0 - 2.8 & 2.8 - 4.0 \\ 
\hline
Statistics         & 0.18      & 0.14      & 0.22      & 0.39      & 0.57      & 0.50 \\ 
Bkg. model         & 0.05      & 0.05      & 0.09      & 0.22      & 0.33      & 0.24 \\ 
Flux               & 0.07      & 0.08      & 0.07      & 0.08      & 0.11      & 0.09 \\ 
Energy scale       & 0.05      & 0.03      & 0.03      & 0.05      & 0.13      & 0.11 \\ 
Kaon int.          & 0.06      & 0.05      & 0.04      & 0.05      & 0.09      & 0.06 \\ 
Sideband tune      & 0.01      & 0.03      & 0.06      & 0.12      & 0.14      & 0.07 \\ 
Sig. model         & 0.02      & 0.00      & 0.04      & 0.12      & 0.17      & 0.12 \\ 
\hline
Total              & 0.22      & 0.18      & 0.27      & 0.49      & 0.72      & 0.59 \\ 
\hline
\end{tabular}
\caption{Fractional statistical and systematic uncertainties reported in bins of $E_{vis}$.}
\label{tab:nusyst}
\end{table}
\endgroup

\begingroup
\squeezetable
\begin{table}[h]
\begin{tabular}{l|cccccc}
\hline
$E_{vis}$ (GeV) & 0.0 - 0.4 & 0.4 - 0.8 & 0.8 - 1.2 & 1.2 - 2.0 & 2.0 - 2.8 & 2.8 - 4.0 \\
\hline
0.0 - 0.4 & 0.821 & 0.361 & -0.078 & -0.133 & -0.069 & -0.046 \\
0.4 - 0.8 & 0.361 & 0.600 & 0.283 & 0.033 & -0.032 & -0.084 \\
0.8 - 1.2 & -0.078 & 0.283 & 0.493 & 0.252 & 0.073 & -0.025 \\
1.2 - 2.0 & -0.133 & 0.033 & 0.252 & 0.342 & 0.196 & 0.081 \\
2.0 - 2.8 & -0.069 & -0.032 & 0.073 & 0.196 & 0.193 & 0.142 \\
2.8 - 4.0 & -0.046 & -0.084 & -0.025 & 0.081 & 0.142 & 0.614 \\
\hline
\end{tabular}
\caption{The statistical covariance introduced by the unfolding procedure on the measurement of $d\sigma/dE_{vis}$, given in units of $10^{-82}$~(cm$^{2}$/nucleon/GeV)$^{2}$.}
\label{tab:nustatcov}
\end{table}
\endgroup

\begingroup
\squeezetable
\begin{table}[h]
\begin{tabular}{l|cccccc}
\hline
$E_{vis}$ (GeV) & 0.0 - 0.4 & 0.4 - 0.8 & 0.8 - 1.2 & 1.2 - 2.0 & 2.0 - 2.8 & 2.8 - 4.0 \\
\hline
0.0 - 0.4 & 0.122 & 0.141 & 0.070 & 0.029 & 0.020 & 0.028 \\
0.4 - 0.8 & 0.141 & 0.170 & 0.087 & 0.036 & 0.022 & 0.028 \\
0.8 - 1.2 & 0.070 & 0.087 & 0.051 & 0.023 & 0.013 & 0.014 \\
1.2 - 2.0 & 0.029 & 0.036 & 0.023 & 0.014 & 0.007 & 0.008 \\
2.0 - 2.8 & 0.020 & 0.022 & 0.013 & 0.007 & 0.007 & 0.008 \\
2.8 - 4.0 & 0.028 & 0.028 & 0.014 & 0.008 & 0.008 & 0.020 \\
\hline
\end{tabular}
\caption{The covariance due to the flux uncertainty on the measurement of $d\sigma/dE_{vis}$, given in units of $10^{-82}$~(cm$^{2}$/nucleon/GeV)$^{2}$.}
\label{tab:nufluxcov}
\end{table}
\endgroup

\begingroup
\squeezetable
\begin{table}[h]
\begin{tabular}{l|cccccc}
\hline
$E_{vis}$ (GeV) & 0.0 - 0.4 & 0.4 - 0.8 & 0.8 - 1.2 & 1.2 - 2.0 & 2.0 - 2.8 & 2.8 - 4.0 \\
\hline
0.0 - 0.4 & 0.202 & 0.178 & 0.073 & 0.015 & -0.011 & 0.018 \\
0.4 - 0.8 & 0.178 & 0.209 & 0.138 & 0.086 & 0.036 & 0.068 \\
0.8 - 1.2 & 0.073 & 0.138 & 0.174 & 0.160 & 0.101 & 0.128 \\
1.2 - 2.0 & 0.015 & 0.086 & 0.160 & 0.187 & 0.137 & 0.171 \\
2.0 - 2.8 & -0.011 & 0.036 & 0.101 & 0.137 & 0.109 & 0.134 \\
2.8 - 4.0 & 0.018 & 0.068 & 0.128 & 0.171 & 0.134 & 0.232 \\
\hline
\end{tabular}
\caption{The covariance due to systematic uncertainties other than the flux, on the measurement of $d\sigma/dE_{vis}$, given in units of $10^{-82}$~(cm$^{2}$/nucleon/GeV)$^{2}$.}
\label{tab:nusyscov}
\end{table}
\endgroup